\documentclass[showpacs,showkeys,amsmath,amssymb,preprint]{revtex4}
\usepackage{epsfig,dcolumn,bm}

\begin{document}

\title{Resonating group method study of kaon-nucleon elastic
scattering in the chiral SU(3) quark model}

\author{F. Huang}
\affiliation{Institute of High Energy Physics, P.O. Box 918-4,
Beijing 100039, PR China \\ Graduate School of the Chinese Academy
of Sciences, Beijing 100039, PR China}
\author{Z.Y. Zhang}
\affiliation{Institute of High Energy Physics, P.O. Box 918-4, Beijing 100039, PR China}
\author{Y.W. Yu}
\affiliation{Institute of High Energy Physics, P.O. Box 918-4, Beijing 100039, PR China}

\begin{abstract}
The chiral SU(3) quark model is extended to include an antiquark
in order to study the kaon-nucleon system. The model input
parameters $b_u$, $m_u$, $m_s$ are taken to be the same as in our
previous work which focused on the nucleon-nucleon and
nucleon-hyperon interactions. The mass of the scalar meson
$\sigma$ is chosen to be 675 MeV and the mixing of $\sigma_0$ and
$\sigma_8$ is considered. Using this model the  kaon-nucleon $S$
and $P$ partial waves phase shifts of isospin $I=0$ and $I=1$ have
been studied by solving a resonating group method (RGM) equation.
The numerical results of $S_{01}$, $S_{11}$, $P_{01}$, $P_{03}$,
and $P_{11}$ partial waves are in good agreement with the
experimental data while the phase shifts of $P_{13}$ partial wave
are a little bit too repulsive when the laboratory momentum of the
kaon meson is greater than 500 MeV in this present calculation.
\end{abstract}

\preprint{nucl-th/0406046}

\maketitle

\section{Introduction}

As is well known, the nonperturbative quantum chromodynamics (QCD)
effect is very important in the light quark system, but up to now
there is no serious practical approach to really solve the
nonperturbative QCD problem. People still need QCD-inspired models
to help. Among these models, the chiral SU(3) quark model
\cite{zyz97} has been successful in reproducing the energies of
the baryon ground states, the binding energy of deuteron, the
nucleon-nucleon ($NN$) scattering phase shifts of different
partial waves, and the hyperon-nucleon ($YN$) cross sections by
the resonating group method (RGM) calculations \cite{zyz97,lrd03}.
In the study of the dibaryon structure, the binding energy of the
$H$ particle obtained from this model is around the threshold of
two $\Lambda$ \cite{pns99}, consistent with the recent
experimental estimation from the binding energy of the double
$\Lambda$ hypernucleus \cite{kim99}. Inspired by these
achievements, we try to extend this model to the systems with
antiquarks to study the baryon-meson interactions. With the
antiquark ($\bar q$) in the meson brought in, the complexity of
the annihilation part in the interactions will appear. As a first
step we start with the study of $KN$ elastic scattering processes
because in the $KN$ system the annihilation to gluons and vacuum
is forbidden and the $u\bar s$ ($d\bar s$) can only annihilate to
kaon mesons.

Another motivation of the present work came from the discovery of
the $\Theta^+$(1540) pentaquark state, an exotic $K^+n$ or $K^0p$
resonance reported by some laboratories recently
\cite{tna03,vvb03,sst03,jba03,aea03,vku04,aai03,aal04}. The
strangeness quantum number of this $\Theta$ particle is $S=+1$ and
the upper limit of the width is about $\Gamma_{\Theta}<25$ MeV. It
may be the first exotic hadron observed and has triggered great
interest and heated discussions. However, the nature of this
particle, its isospin, parity, and angular momentum, is still
going to be determined. In order to obtain a reasonable
interpretation of the data of the $uudd\bar{s}$ system, a prior
understanding of the kaon-nucleon interaction on a quark level is
important and necessary.

Actually, the $KN$ scattering had aroused particular interest in
the past due to the kaon meson's high penetrating power
\cite{wrc85,yab87}, which makes the kaon one of the deepest probes
of the nuclear medium in the energy range between 0 and 1 GeV/$c$.
The model based on hadronic degrees of freedom \cite{rbu90} can
give a good description of $KN$ interaction, but Buettgen {\it et
al.} had to add the exchange of a short range ($\sim$ 0.2 fm)
repulsive scalar meson in order to reproduce the $S$-wave phase
shifts in the isospin $I=0$ channel. The range of this repulsion
is smaller than the nucleon radius, which clearly shows that the
quark substructure of the kaon mesons and nucleons cannot be
neglected. In Ref. \cite{sle03}, the $KN$ phase shifts are
calculated within a constituent quark model by numerically solving
the RGM equation. In that calculation, the quark-quark potential
includes gluon, pion, and sigma exchanges and the ground state
energies of mesons can be reproduced, but the agreement of the
obtained results with the experimental phase shifts is quite poor.
Recently, Wang {\it et al.} \cite{hjw03} gave a study on the $KN$
elastic scattering in a quark potential model. Their results are
consistent with the experimental data, but in their model, a
factor of color octet component is added arbitrarily and the size
parameter of harmonic oscillator is chosen to be $b_u=0.255$ fm,
which is too small compared with the radius of nucleon.

The goal of the present work aims at studying the $KN$ elastic
scattering phase shifts of $S$ and $P$ partial waves of isospin
$I=0$ and $I=1$ in the framework of the chiral SU(3) quark model
by carrying on a resonating group method calculation. We take the
same input model parameters $b_u$, $m_u$, $m_s$ as in our previous
work \cite{zyz97, lrd03}, which successfully explained the
existing $NN$ and $YN$ experimental data. The difference is that
in the present work the mass of the scalar meson $\sigma$ is
chosen to be 675 MeV (in our previous work $m_\sigma=$595 or 625
MeV) and the mixing between $\sigma_0$ and $\sigma_8$ is
considered. By this means the attraction of $\sigma$ meson in $KN$
$S_{01}$ partial wave can be reduced a lot. Except for the case of
$P_{13}$, the numerical results of different partial waves are in
agreement with the experimental data. In comparison with the
previous results \cite{sle03,nbl02}, our calculation achieves a
considerable improvement on the theoretical phase shifts. In this
sense, it means that our model also works well when an anti-quark
is added in the system (at least for the $KN$ system), so that one
can regard that the interactions between two quarks obtained from
this model is almost reasonable, which is useful for studying the
structure of the $\Theta^+$(1540) pentaquark state from the
constituent quark model point of view.

The paper is organized as follows. In the next section the
framework of the chiral SU(3) quark model and the RGM approach
applying to the $KN$ system are briefly introduced. The calculated
results of the isospin $I=0$ and $I=1$ $KN$ phase shifts of $S$,
$P$ partial waves are shown in Sec. III, as well as some
discussions are made in this section. Finally, conclusions are
drawn in Sec. IV.

\section{Formulation}

\subsection{The model}

Following Georgi's idea \cite{hge84}, the interaction Lagrangian
of the quark-chiral SU(3) field can be written as
\begin{eqnarray}
{\cal L}_{I} = -g_{ch} (\bar{\psi}_{L}\Sigma\psi_{R} +
\bar{\psi}_{R}\Sigma^{+}\psi_{L}),
\end{eqnarray}
with $g_{ch}$ being the quark-chiral-field coupling constant,
$\psi_{L}$ and $\psi_{R}$ being the quark-left and quark-right
spinors, respectively, and
\begin{eqnarray}
\Sigma =\mbox{exp}{[i \pi_{a} \lambda_{a}/f]}, ~~~~ a=1,2,...,8.
\end{eqnarray}
where $\pi_{a}$ is the Goldstone boson field and $\lambda_{a}$ the
Gell-Mann matrix of the flavor SU(3) group. Generalizing the
linear realization of $\Sigma$ from the SU(2) case to the SU(3)
case, one obtains
\begin{eqnarray}
\Sigma = \sum^{8}_{a=0} \sigma_{a} \lambda_{a} + i \sum^{8}_{a=0}
\pi_a \lambda_a,
\end{eqnarray}
and the interaction Lagrangian
\begin{eqnarray}
{\cal L}_{I} = -g_{ch} \bar{\psi} \left( \sum^{8}_{a=0} \sigma_a
\lambda_a + i \sum^{8}_{a=0} \pi_a \lambda_a \gamma_5 \right)
\psi,
\end{eqnarray}
where $\lambda_{0}$ is a unitary matrix,
$\sigma_{0},...,\sigma_{8}$ are the scalar nonet fields, and
$\pi_{0},...,\pi_{8}$ the pseudoscalar nonet fields. Clearly,
${\cal L}_{I}$ is invariant under the infinitesimal chiral
SU(3)$_{L} \times$ SU(3)$_{R}$ transformation. Consequently, one
obtains the interactive Hamiltonian as
\begin{eqnarray}\label{hami}
H_{ch} = g_{ch} F(\bm q^{2}) \bar{\psi} \left( \sum^{8}_{a=0}
\sigma_a \lambda_a + i \sum^{8}_{a=0} \pi_a \lambda_a \gamma_5
\right) \psi.
\end{eqnarray}
Here we insert a form factor $F(\bm q^{2})$ to describe the
chiral-field structure \cite{ito90,amk91}. As usual, $F(\bm
q^{2})$ is taken as
\begin{eqnarray}\label{faca}
F(\bm q^{2})=\left(\frac{\Lambda^2}{\Lambda^2+\bm
q^2}\right)^{1/2},
\end{eqnarray}
and the cutoff mass $\Lambda$ indicates the chiral symmetry
breaking scale \cite{ito90,amk91,abu91,emh91}.

Form Eqs. (\ref{hami}) and (\ref{faca}) the SU(3)
chiral-field-induced quark-quark potentials can be derived, and
their expressions are given in the following:
\begin{eqnarray}
V_{\sigma_a}({\bm r}_{ij})=-C(g_{ch},m_{\sigma_a},\Lambda)
X_1(m_{\sigma_a},\Lambda,r_{ij}) [\lambda_a(i)\lambda_a(j)] +
V_{\sigma_a}^{\bm {l \cdot s}}({\bm r}_{ij}),
\end{eqnarray}
\begin{eqnarray}
V_{\sigma_a}^{\bm {l \cdot s}}({\bm r}_{ij})&=&
-C(g_{ch},m_{\sigma_a},\Lambda)\frac{m^2_{\sigma_a}}{4m_{q_i}m_{q_j}}
\left\{G(m_{\sigma_a}r_{ij})-\left(\frac{\Lambda}{m_{\sigma_a}}\right)^3
G(\Lambda r_{ij})\right\} \nonumber\\
&&\times[{\bm L \cdot ({\bm \sigma}_i+{\bm
\sigma}_j)}][\lambda_a(i)\lambda_a(j)],
\end{eqnarray}
and
\begin{eqnarray}
V_{\pi_a}({\bm r}_{ij})=C(g_{ch},m_{\pi_a},\Lambda)
\frac{m^2_{\pi_a}}{12m_{q_i}m_{q_j}} X_2(m_{\pi_a},\Lambda,r_{ij})
({\bm \sigma}_i\cdot{\bm \sigma}_j)[\lambda_a(i)\lambda_a(j)],
\end{eqnarray}
with
\begin{eqnarray}
C(g_{ch},m,\Lambda)=\frac{g^2_{ch}}{4\pi}
\frac{\Lambda^2}{\Lambda^2-m^2} m,
\end{eqnarray}
\begin{eqnarray}
\label{x1mlr} X_1(m,\Lambda,r)=Y(mr)-\frac{\Lambda}{m} Y(\Lambda
r),
\end{eqnarray}
\begin{eqnarray}
X_2(m,\Lambda,r)=Y(mr)-\left(\frac{\Lambda}{m}\right)^3 Y(\Lambda
r),
\end{eqnarray}
\begin{eqnarray}
Y(x)=\frac{1}{x}e^{-x},
\end{eqnarray}
\begin{eqnarray}
G(x)=\frac{1}{x}\left(1+\frac{1}{x}\right)Y(x),
\end{eqnarray}
and $m_{\sigma_a}$ being the mass of the scalar meson and
$m_{\pi_a}$ the mass of the pseudoscalar meson.

As mentioned in Ref. \cite{zyz94}, in the chiral SU(3) quark model
the interaction induced by the coupling of chiral field describes
the nonperturbative QCD effect of the low-momentum medium-distance
range. To study the hadron structure and hadron-hadron dynamics,
one still needs an effective one-gluon-exchange interaction
$V^{OGE}_{ij}$ which governs the short-range perturbative QCD
behavior,
\begin{eqnarray}
V^{OGE}_{ij}&=&\frac{1}{4}g_{i}g_{j}\left(\lambda^c_i\cdot\lambda^c_j\right)
\left\{\frac{1}{r_{ij}}-\frac{\pi}{2} \delta({\bm r}_{ij})
\left(\frac{1}{m^2_{q_i}}+\frac{1}{m^2_{q_j}}+\frac{4}{3}\frac{1}{m_{q_i}m_{q_j}}
({\bm \sigma}_i \cdot {\bm \sigma}_j)\right)\right\} \nonumber \\
&&+V_{OGE}^{\bm l \cdot \bm s},
\end{eqnarray}
with
\begin{eqnarray}
V_{OGE}^{\bm l \cdot \bm
s}=-\frac{1}{16}g_ig_j\left(\lambda^c_i\cdot\lambda^c_j\right)
\frac{3}{m_{q_i}m_{q_j}}\frac{1}{r^3_{ij}}{\bm L \cdot ({\bm
\sigma}_i+{\bm \sigma}_j)},
\end{eqnarray}
and a confinement potential $V^{conf}_{ij}$ which provides the
nonperturbative QCD effect in the long distance,
\begin{eqnarray}
V_{ij}^{conf}=-a_{ij}^{c}(\lambda_{i}^{c}\cdot\lambda_{j}^{c})r_{ij}^2
-a_{ij}^{c0}(\lambda_{i}^{c}\cdot\lambda_{j}^{c}).
\end{eqnarray}
For the $KN$ system, we have to extend our chiral SU(3) quark
model to the case with an antiquark. Now, the total Hamiltonian of
$KN$ system is written as
\begin{eqnarray}
\label{hami5q}
H=\sum_{i=1}^{5}T_{i}-T_{G}+\sum_{i<j=1}^{4}V_{ij}+\sum_{i=1}^{4}V_{i\bar
5},
\end{eqnarray}
where $T_G$ is the kinetic energy operator of the center of mass
motion, and $V_{ij}$ and $V_{i\bar 5}$ represent the interactions
between quark-quark ($qq$) and quark-antiquark ($q{\bar q}$),
respectively,
\begin{eqnarray}
V_{ij}= V^{OGE}_{ij} + V^{conf}_{ij} + V^{ch}_{ij},
\end{eqnarray}
\begin{eqnarray}
V^{ch}_{ij}=\sum^{8}_{a=0}V_{\sigma_a}(\bm
r_{ij})+\sum^{8}_{a=0}V_{\pi_a} (\bm r_{ij}).
\end{eqnarray}
The interaction between $u (d)$ and $\bar s$ includes two parts
\cite{fhu03}: direct interaction and annihilation parts,
\begin{eqnarray}
V_{i\bar 5}=V^{dir}_{i\bar 5}+V^{ann}_{i\bar 5},
\end{eqnarray}
with
\begin{eqnarray}
V_{i\bar 5}^{dir}=V_{i\bar 5}^{conf}+V_{i\bar 5}^{OGE}+V_{i\bar
5}^{ch},
\end{eqnarray}
and
\begin{eqnarray}
V_{i\bar
5}^{conf}=-a_{i5}^{c}\left(-\lambda_{i}^{c}\cdot{\lambda_{5}^{c}}^*\right)r_{i5}^2
-a_{i5}^{c0}\left(-\lambda_{i}^{c}\cdot{\lambda_{5}^{c}}^*\right),
\end{eqnarray}
\begin{eqnarray}
V^{OGE}_{i\bar 5}&=&\frac{1}{4}g_{i}g_{s}\left(-\lambda^c_i\cdot{\lambda^c_5}^*\right)
\left\{\frac{1}{r_{i5}}-\frac{\pi}{2} \delta({\bm r}_{i5})
\left(\frac{1}{m^2_{q_i}}+\frac{1}{m^2_{s}}+\frac{4}{3}\frac{1}{m_{q_i}m_{s}}
({\bm \sigma}_i \cdot {\bm \sigma}_5)\right)\right\}  \nonumber \\
&&-\frac{1}{16}g_ig_s\left(-\lambda^c_i\cdot{\lambda^c_5}^*\right)
\frac{3}{m_{q_i}m_{q_5}}\frac{1}{r^3_{i5}}{\bm L \cdot ({\bm
\sigma}_i+{\bm \sigma}_5)},
\end{eqnarray}
\begin{eqnarray}
V_{i\bar{5}}^{ch}=\sum_{j}(-1)^{G_j}V_{i5}^{ch,j}.
\end{eqnarray}
Here $(-1)^{G_j}$ describes the G parity of the $j$th meson. For
the $KN$ system, $u(d)\bar{s}$ can only annihilate into a $K$
meson, i.e.,
\begin{eqnarray}
V_{i\bar 5}^{ann}=V_{ann}^{K},
\end{eqnarray}
with
\begin{eqnarray}
V_{ann}^{K}&=&C^K_{ann}\left(\frac{1-{\bm \sigma}_q \cdot {\bm
\sigma}_{\bar{q}}}{2}\right)_{spin}\left(\frac{2 + 3\lambda_q
\cdot \lambda^*_{\bar{q}}}{6}\right)_{color}
\left(\frac{38+3\lambda_q \cdot \lambda^*_{\bar
q}}{18}\right)_{flavor} \frac{\Lambda^2}{r}\mbox{e}^{-\Lambda r},
\end{eqnarray}
\begin{equation}
C^K_{ann}=-\frac{\tilde{g}_{ch}^2}{4\pi}\frac{1}{m_K^2-(\tilde{m}+\tilde{m}_{s})^2},
\end{equation}
where $\tilde{g}_{ch}$ is the effective coupling constant of
chiral field in the annihilation case and $\tilde{m}$ represents
the effective quark mass. Actually, $\tilde{m}$ is quark momentum
dependent; here we treat it as an effective mass. In the present
form of the annihilation interaction $V^K_{ann}$, a form factor
$F(\bm q^2)$ [Eq. (\ref{faca})], which is also used in the vertex
of the quark-chiral-field coupling, is inserted to flat the sharp
behavior of the $\delta$ function. In this work we treat
$C^K_{ann}$ as a parameter and adjust it to fit the mass of kaon
meson.

\subsection{Determination of parameters}

We have three initial input parameters: the harmonic-oscillator
width parameter $b_u$, the up (down) quark mass $m_{u(d)}$, and
the strange quark mass $m_s$. These three parameters are taken to
be the same as in our previous work \cite{zyz97, lrd03}, i.e.,
$b_u=0.5$ fm, $m_{u(d)}=313$ MeV, and $m_s=470$ MeV. By some
special constraints, the other model parameters are fixed in the
following way: the chiral coupling constant $g_{ch}$ is fixed by
\begin{eqnarray}
\frac{g^{2}_{ch}}{4\pi} = \left( \frac{3}{5} \right)^{2}
\frac{g^{2}_{NN\pi}}{4\pi} \frac{m^{2}_{u}}{M^{2}_{N}},
\end{eqnarray}
with $g^{2}_{NN\pi}/4\pi=13.67$ taken as the experimental value.
The masses of the mesons are also adopted to the experimental
values, except for the $\sigma$ meson, where its mass is treated
as an adjustable parameter; in this work, it is adopted to be 675
MeV. The cutoff radius $\Lambda^{-1}$ is taken to be the value
close to the chiral symmetry breaking scale
\cite{ito90,amk91,abu91,emh91}. After the parameters of chiral
fields are fixed, the one-gluon-exchange coupling constants
$g_{u}$ and $g_{s}$ can be determined by the mass splits between
$N$, $\Delta$ and $\Sigma$, $\Lambda$, respectively. The
confinement strengths $a^{c}_{uu}$, $a^{c}_{us}$ and $a^{c}_{ss}$
are fixed by the stability conditions of $N$, $\Lambda$, and
$\Xi$, and the zero point energies $a^{c0}_{uu}$, $a^{c0}_{us}$,
and $a^{c0}_{ss}$ by fitting the masses of $N$, $\Sigma$ and
$\overline{\Xi+\Omega}$, respectively. About $C^K_{ann}$, we
adjust it to fit the mass of kaon meson. The resultant model
parameters are tabulated in Table \ref{para} and the masses of
octet and decuplet baryons obtained from this set of parameters
are listed in Table \ref{baryonmass}.

{\small
\begin{table}[htb]
\caption{\label{para} Model parameters. The meson masses and the
cutoff masses: $m_{\sigma'}=980$ MeV, $m_{\kappa}=1430$ MeV,
$m_{\epsilon}=980$ MeV, $m_{\sigma}=675$ MeV, $m_{\pi}=138$ MeV,
$m_K=495$ MeV, $m_{\eta}=549$ MeV, $m_{\eta'}=957$ MeV,
$\Lambda=1500$ MeV for $\kappa$ and 1100 MeV for other mesons.}
\begin{tabular*}{165mm}{@{\extracolsep\fill}ccccc}
\hline\hline
 & $m_u$ (MeV) && 313 &  \\
 & $m_s$ (MeV) && 470 &  \\
 & $b_u$ (fm)  && 0.5 &  \\
 & $g_u$     && 0.886  & \\
 & $g_s$     && 0.755  & \\
 & $a^c_{uu}$ (MeV/fm$^2$) && 52.40 & \\
 & $a^c_{us}$ (MeV/fm$^2$) && 75.30 & \\
 & $a^{c0}_{uu}$ (MeV)  && $-$50.37 & \\
 & $a^{c0}_{us}$ (MeV)  && $-$66.80 & \\
 & $C^K_{ann}$ (fm$^2$) && $-$0.137 & \\
\hline\hline
\end{tabular*}
\end{table}}

{\small
\begin{table}[htb]
\caption{\label{baryonmass} The masses of octet and decuplet
baryons.}
\begin{tabular*}{165mm}{@{\extracolsep\fill}lcccccccc}
\hline\hline
       & $N$ & $\Sigma$ & $\Xi$ & $\Lambda$ & $\Delta$ & $\Sigma^\ast$ & $\Xi^\ast$ & $\Omega$  \\
\hline
Theor. & 939 &   1194   & 1334  &   1116    &    1237  &       1375    &    1515    &   1657    \\
Expt.  & 939 &   1194   & 1319  &   1116    &    1237  &       1385    &    1530    &   1672    \\
\hline\hline
\end{tabular*}
\end{table}}

In our calculation, the meson mixing between the flavor singlet
and octet mesons is considered, i.e., $\eta$, $\eta'$ mesons are
mixed by $\eta_0$, $\eta_8$:
\begin{equation}
\eta'=\eta_8 \sin\theta^{PS}+\eta_0 \cos\theta^{PS},\nonumber
\end{equation}
\begin{equation}
\eta=\eta_8 \cos\theta^{PS}-\eta_0 \sin\theta^{PS},
\end{equation}
with the mixing angle $\theta^{PS}$ taken to be the usual value
$-23^\circ$ and  $\sigma$, $\epsilon$ mesons are ideally mixed by
$\sigma_0$, $\sigma_8$:
\begin{equation}
\sigma=\sigma_8 \sin\theta^S+\sigma_0 \cos\theta^S, \nonumber
\end{equation}
\begin{equation}
\epsilon=\sigma_8 \cos\theta^S-\sigma_0 \sin\theta^S,
\end{equation}
with $\theta^S=35.264^\circ$, which means that $\sigma$ only acts
on the $u(d)$ quark, and $\epsilon$ on the $s$ quark,
respectively. Under this ideal mixing, the scalar meson exchange
interactions between $u(d)$ and $\bar{s}$ are totally vanished, so
that the attraction force of scalar meson between $K$ and $N$ can
be reduced a lot.

\subsection{The RGM approach applying to the $KN$ system}

In this section, we present the applying of the resonating group
method (RGM) to the $KN$ system. We take the following choice of
the coordinates to construct the total wave function of the
system:
\begin{equation}
{\bm \xi}_1={\bm r}_2-{\bm r}_1,
\end{equation}
\begin{equation}
{\bm \xi}_2={\bm r}_3-\frac{{\bm r}_1+{\bm r}_2}{2},
\end{equation}
\begin{equation}
{\bm \xi}_3={\bm r}_5-{\bm r}_4,
\end{equation}
\begin{equation}
{\bm R}_{KN}=\frac{{\bm r}_1+{\bm r}_2+{\bm r}_3}{3}-\frac{m_u
{\bm r}_4+m_s {\bm r}_5}{m_u+m_s},
\end{equation}
\begin{equation}
\bm R_{c.m.}=\frac{m_u (\bm r_1+\bm r_2+\bm r_3+\bm r_4)+m_s \bm
r_5}{4m_u+m_s}.
\end{equation}
Here, $\bm r_i$ is the coordinate of the $i$th quark, ${\bm
\xi}_1$ and ${\bm \xi}_2$ are the internal coordinates for the
cluster $N$, and ${\bm \xi}_3$ the internal coordinate for $K$.
${\bm R}_{KN}$ is the relative coordinate between $K$ and $N$, and
$\bm R_{c.m.}$ is the center of mass coordinate of the total
system.

Following the cluster model calculation \cite{mka77,mok81,ust88},
the RGM wave function is written as
\begin{eqnarray}
\Psi={\cal A}[{\hat \phi}_N(\bm \xi_1,\bm \xi_2 ) {\hat
\phi}_K(\bm \xi_3) \chi_{rel}({\bm R}_{KN})Z(\bm R_{c.m.})]_{ST},
\end{eqnarray}
with
\begin{equation}
\phi_N(\bm \xi_1,\bm \xi_2)=\left(\frac{m_u
\omega}{2\pi}\right)^{3/4} \left(\frac{2m_u
\omega}{3\pi}\right)^{3/4} \mbox{exp}\left[-m_u
\omega\left(\frac{\bm \xi_1^2}{4}+\frac{\bm
\xi_2^2}{3}\right)\right],
\end{equation}
\begin{equation}
\phi_K(\bm \xi_3)=\left(\frac{\omega}{\pi}\frac{m_u
m_s}{m_u+m_s}\right)^{3/4}
\mbox{exp}\left[-\frac{\omega}{2}\frac{m_u m_s}{m_u+m_s}\bm
\xi_3^2\right],
\end{equation}
\begin{equation}
Z(\bm R_{c.m.})=\left(\frac{\omega}{\pi}(4m_u+m_s)\right)^{3/4}
\mbox{exp}\left[-\frac{\omega}{2}(4m_u+m_s)\bm R_{c.m.}^2\right].
\end{equation}
Here $\phi_N(\bm \xi_1,\bm \xi_2)$ and $\phi_K(\bm \xi_3)$ denote
the internal wave function in coordinate space of cluster $N$ and
$K$, respectively. ${\hat \phi}_N\left(\bm \xi_1,\bm \xi_2
\right)$ represents the antisymmetrized wave function of cluster
$N$ and ${\hat \phi}_K\left(\bm \xi_3\right)$, the wave function
of cluster $K$ with $N$ and $K$ further specifying all the quantum
numbers of the relevant cluster. $\chi_{rel}({\bm R}_{KN})$ is the
trial wave function of the relative motion between interacting
clusters $K$ and $N$, and $Z(\bm R_{c.m.})$ is the wave function
of the motion of the total center of mass. The oscillator
frequency $\omega$ is associated with the width parameter $b_u$ by
the constituent quark mass $m_u$:
\begin{eqnarray}
\frac{1}{b_i^2}=m_i\omega.
\end{eqnarray}
The symbol $\cal A$ is the antisymmetrizing operator defined as
\begin{equation}
{\cal A}\equiv{1-\sum_{i \in N}P_{i4}}\equiv{1-3P_{34}}.
\end{equation}
$S$ and $T$ denote the total spin and isospin of the $KN$ system,
respectively. Substituting $\Psi$ into the projection equation
\begin{equation}
\langle \delta\Psi|(H-E)|\Psi \rangle=0,
\end{equation}
where
\begin{eqnarray}
E=E_K+E_N+E_{rel},
\end{eqnarray}
with $E$, $E_{K}$, $E_{N}$, and $E_{rel}$ being the total energy,
the inner energies of clusters $K$ and $N$, and the relative
energy between clusters $K$ and $N$, respectively, we obtain RGM
equation
\begin{eqnarray}
\int {\cal L}(\bm R', \bm R)\chi_{rel}(\bm R) d\bm R =0,
\end{eqnarray}
with
\begin{eqnarray}
{\cal L}(\bm R', \bm R)={\cal H}(\bm R', \bm R)-E{\cal N}(\bm R',
\bm R),
\end{eqnarray}
where the Hamiltonian kernel $\cal H$ and normalization kernel
$\cal N$ can, respectively, be calculated by
\begin{eqnarray}
\left\{
       \begin{array}{c}
          {\cal H}(\bm R', \bm R)\\
          {\cal N}(\bm R', \bm R)
       \end{array}
\right\}
=\left<[{\hat \phi}_N(\bm \xi_1,\bm \xi_2 )
{\hat \phi}_K(\bm \xi_3) \delta(\bm R'-{\bm
R}_{KN})Z(\bm R_{c.m.})]_{ST}\right.\left| \left\{
\begin{array}{c}
          H \\
          1
       \end{array}
\right\}
\right| \nonumber \\
\left.{\cal A}[{\hat \phi}_N(\bm \xi_1,\bm \xi_2)
{\hat \phi}_K(\bm \xi_3\delta(\bm R-{\bm R}_{KN})Z(\bm R_{c.m.})_{ST}\right>.
\end{eqnarray}

In the actual calculation, the unknown $\chi_{rel}$ is determined
in the following way: First, we perform a partial wave expansion,
\begin{equation}
\chi_{rel}({\bm R}_{KN})=\sum_L\chi_{rel}^L({\bm R}_{KN}),
\end{equation}
and then, for a bound-state problem, $\chi_{rel}^L({\bm R}_{KN})$
is expanded as
\begin{eqnarray}\label{expan}
\chi_{rel}^L({\bm R}_{KN})&=&\sum_{i=1}^n c_i \int
\left(\frac{\omega \mu_{KN}}{\pi}\right)^{3/4}
\mbox{exp}\left[-\frac{\omega \mu_{KN}}{2}({\bm R}_{KN}-\bm
S_i)^2\right]
Y_{LM}(\hat {\bm S}_i)\mbox{d}\hat{\bm S}_i  \nonumber \\
&=&\sum_{i=1}^n c_i \frac{1}{R_{KN}}u^L(R_{KN}, S_i)Y_{LM}({\hat
{\bm R}}_{KN}),
\end{eqnarray}
with
\begin{eqnarray}
u^L(R_{KN}, S_i)&\equiv&4\pi
R_{KN}\left(\frac{\omega\mu_{KN}}{\pi}\right)^{3/4}
\mbox{exp}\left[-\frac{1}{2}\omega\mu_{KN}(R_{KN}^2+S_i^2)\right] \nonumber \\
&&\times i_L(\omega \mu_{KN}R_{KN}S_i),
\end{eqnarray}
where $S_i$ is called the generate coordinate, $\mu_{KN}$ is the
reduced mass of $KN$ system, and $i_L$ the $L$th modified
spherical Bessel function. Usually $\chi_{rel}({\bm R}_{KN})$ is
also expanded as
\begin{equation}
\chi_{rel}({\bm
R}_{KN})=\sum_L\frac{1}{R_{KN}}\chi_{rel}^L(R_{KN})Y_{LM}({\hat
{\bm R}}_{KN}),
\end{equation}
so equivalently, Eq. (\ref{expan}) can be written in a compact
form
\begin{equation}
\chi_{rel}^L(R_{KN})=\sum_{i=1}^n c_i u^L(R_{KN}, S_i).
\end{equation}
Now all the information about the relative wave function is
contained in the coefficients $c_i^,s$ which are left to be
solved. Performing variational procedure, one can deduce a $L$th
partial-wave equation for the bound-state problem,
\begin{eqnarray}\label{eqbound}
\sum_{j=1}^n {\cal L}^L_{ij}c_j=0~~~(i=1,...,n),
\end{eqnarray}
with
\begin{equation}
{\cal L}^L_{ij}=\int u^L(R', S_i){\cal L}^L(R', R) u^L(R,
S_j)R'RdR'dR,
\end{equation}
\begin{equation}\label{rgmkernel}
{\cal L}^L(R', R)=\int Y^*_{LM}(\hat {\bm R'}){\cal L}(\bm R', \bm
R) Y_{LM}(\hat {\bm R})d\hat {\bm R'}d\hat {\bm R}.
\end{equation}
Solving Eq. $(\ref{eqbound})$, we can get the binding energy and
the corresponding wave function of the two-cluster system.

For a scattering problem, the relative wave function is expanded
as
\begin{equation}
\chi_{rel}^L(R_{KN})=\sum_{i=1}^n c_i {\tilde u}^L(R_{KN}, S_i),
\end{equation}
\begin{equation}
\tilde{u}^L(R_{KN}, S_i)\equiv\left\{
\begin{array}{lr}
          p_i u^L(R_{KN}, S_i),~~&R_{KN}\leq R_C \\
          \left[h_L^-(k_{KN}R_{KN})-s_i h_L^+(k_{KN}R_{KN})\right]R_{KN},~~&R_{KN}\geq R_C
\end{array} \right.
\end{equation}
with $h_L^{\pm}$ being $L$th spherical Hankel functions,
$k_{KN}=\sqrt{2\mu_{KN}E_{rel}}$ the momentum of relative motion,
and $R_C$ a cutoff radius beyond which all the strong interactions
can be disregarded. The complex parameters $p_i$ and $s_i$ are
determined by the smoothness condition at $R_{KN}=R_C$ and
${c_i}^,s$ satisfy $\sum_{i=1}^nc_i=1$. Performing variational
procedure, a $L$th partial-wave equation for the scattering
problem can be deduced as
\begin{eqnarray}\label{eqscatter}
\sum_{j=1}^{n-1} {\tilde {\cal L}}^L_{ij}c_j={\tilde {\cal
M}}_i^L~~~(i=1,...,n),
\end{eqnarray}
with
\begin{equation}
{\tilde {\cal L}}^L_{ij}={\tilde {\cal K}}^L_{ij}-{\tilde {\cal
K}}^L_{in}-{\tilde {\cal K}}^L_{nj}+{\tilde {\cal K}}^L_{nn},
\end{equation}
\begin{equation}
{\tilde {\cal M}}^L_i={\tilde {\cal K}}^L_{nn}-{\tilde {\cal
K}}^L_{in},
\end{equation}
and
\begin{equation}
{\tilde {\cal K}}^L_{ij}=\int {\tilde u}^L(R', S_i){\cal L}^L(R',
R) {\tilde u}^L(R, S_j)R'RdR'dR,
\end{equation}
where the RGM kernel ${\cal L}^L(R', R)$ is defined in Eq.
$(\ref{rgmkernel})$. Before solving the Eq. $(\ref{eqscatter})$,
we have to calculate the kernel ${\tilde {\cal K}}^L_{ij}$.
Considering the asymptotic form of spherical Hankel functions,
${\tilde {\cal K}}^L_{ij}$ can be written as
\begin{eqnarray}
{\tilde {\cal K}}^L_{ij}=p_ip_j({\cal L}^L_{ij}-K^{L(ex)}_{ij}),
\end{eqnarray}
\begin{eqnarray}
K_{ij}^{L(ex)}=\int^\infty_{R_C} u^L(R, S_i)
\left(-\frac{\hbar^2}{2\mu_{KN}}\frac{d^2}{dR^2}+\frac{\hbar^2}{2\mu_{KN}}\frac{L(L+1)}{R^2}-E_{rel}\right)
u^L(R, S_j)dR.
\end{eqnarray}
Having solved Eq. $(\ref{eqscatter})$, the $S$-matrix element
$S^L$ and the phase shifts $\delta_L$ are given by
\begin{eqnarray}
S^L\equiv e^{2i\delta_L}=\sum_{i=1}^n c_i s_i.
\end{eqnarray}

\section{Results of $KN$ phase shifts and discussions}

\begin{figure}[h]
\vglue 2.3cm
\epsfig{file=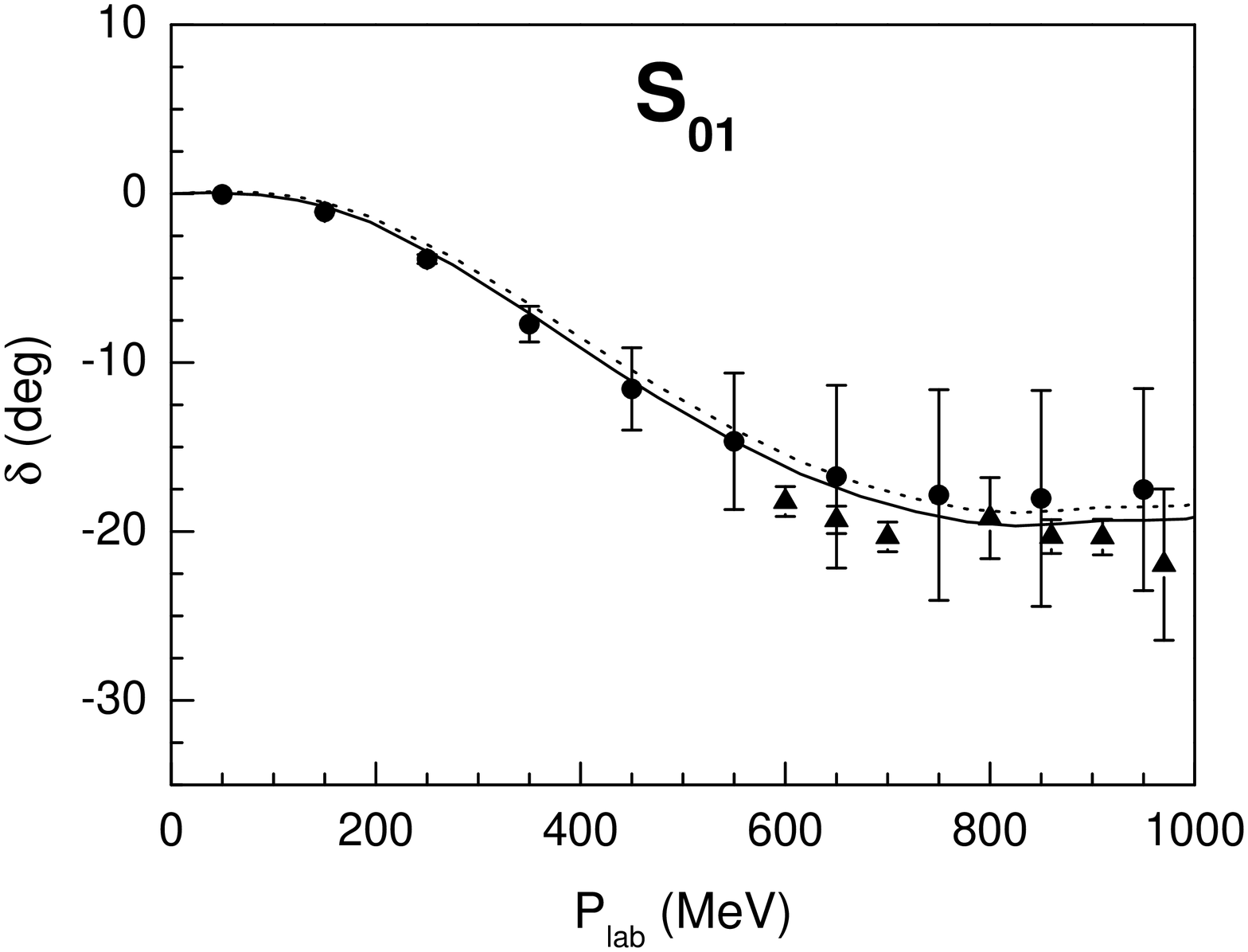,width=7.7cm}
\epsfig{file=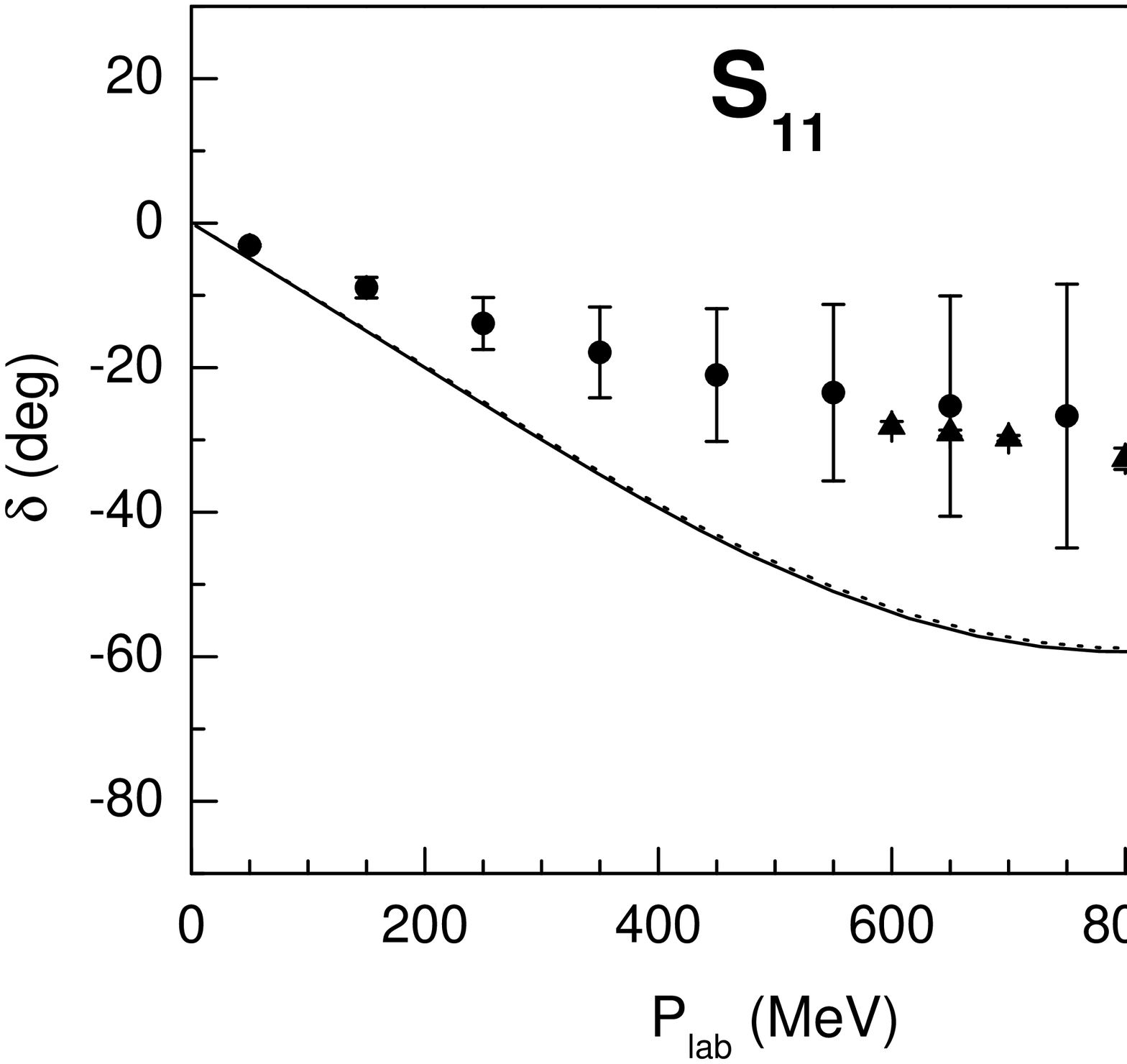,width=7.7cm}
\vglue -2.3cm
\caption{\small
\label{s01s11} $KN$ S-wave phase shifts as a function of the
laboratory momentum of kaon meson. The solid lines represent the
results obtained by considering $\theta^S=35.264^\circ$ while the
dotted lines $\theta^S=-18^\circ$. The hole circles and the
triangles correspond respectively to the phase shifts analysis of
Hyslop {\it et al.} \cite{jsh92} and Hashimoto \cite{kha84}.}
\end{figure}

\begin{figure}[h]
\vglue 2.3cm \epsfig{file=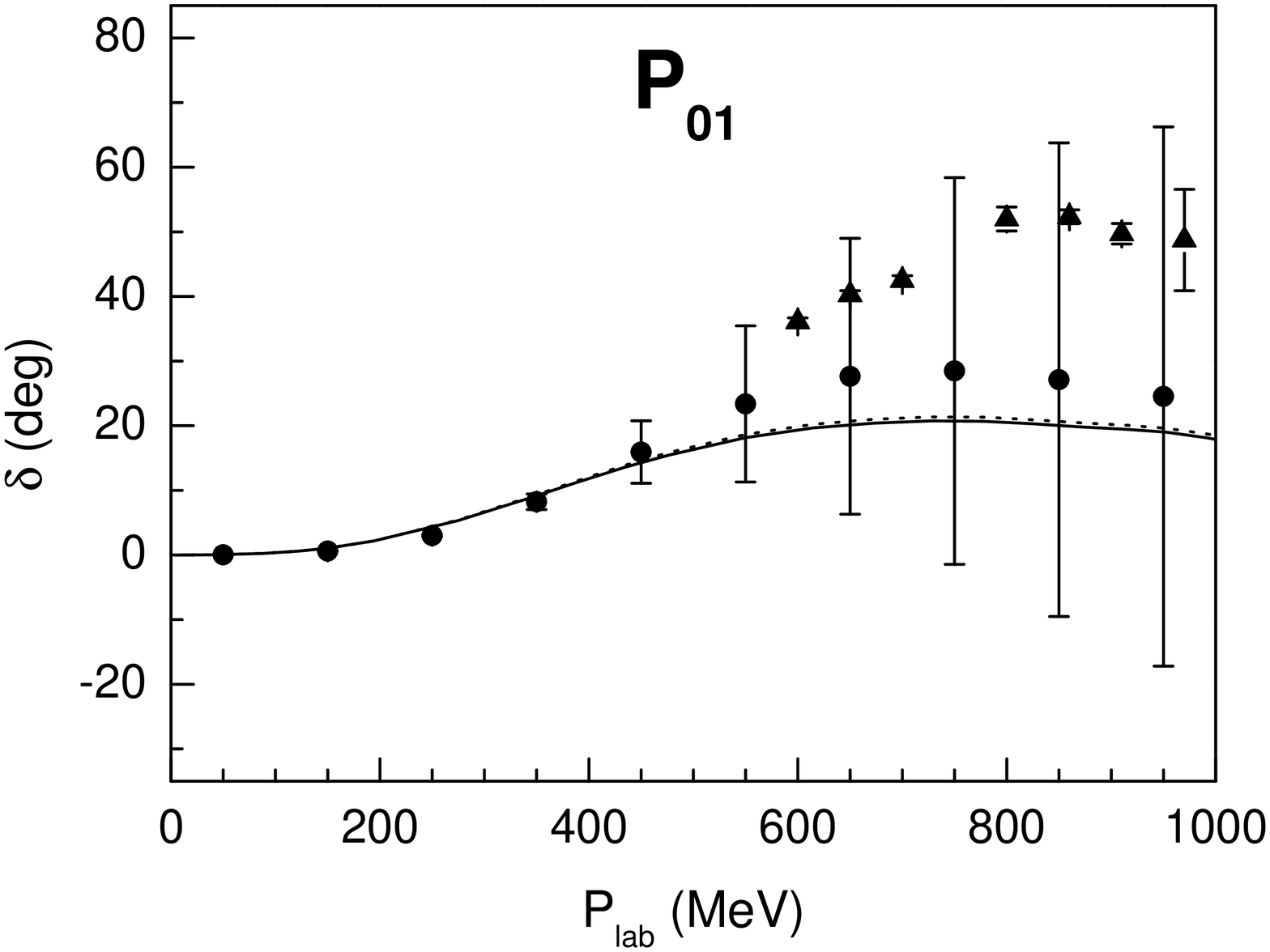,width=7.7cm}
\epsfig{file=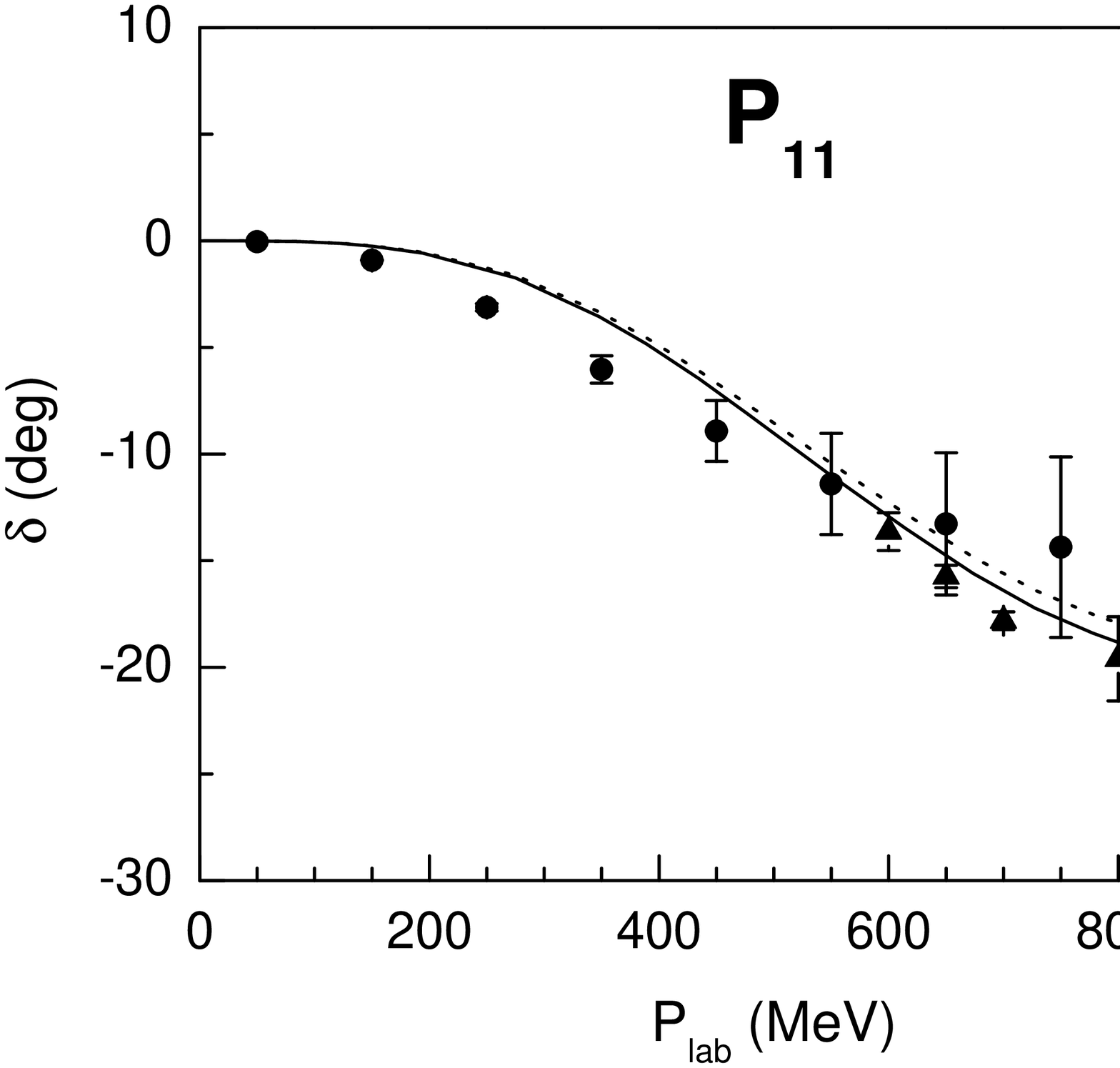,width=7.7cm}
\epsfig{file=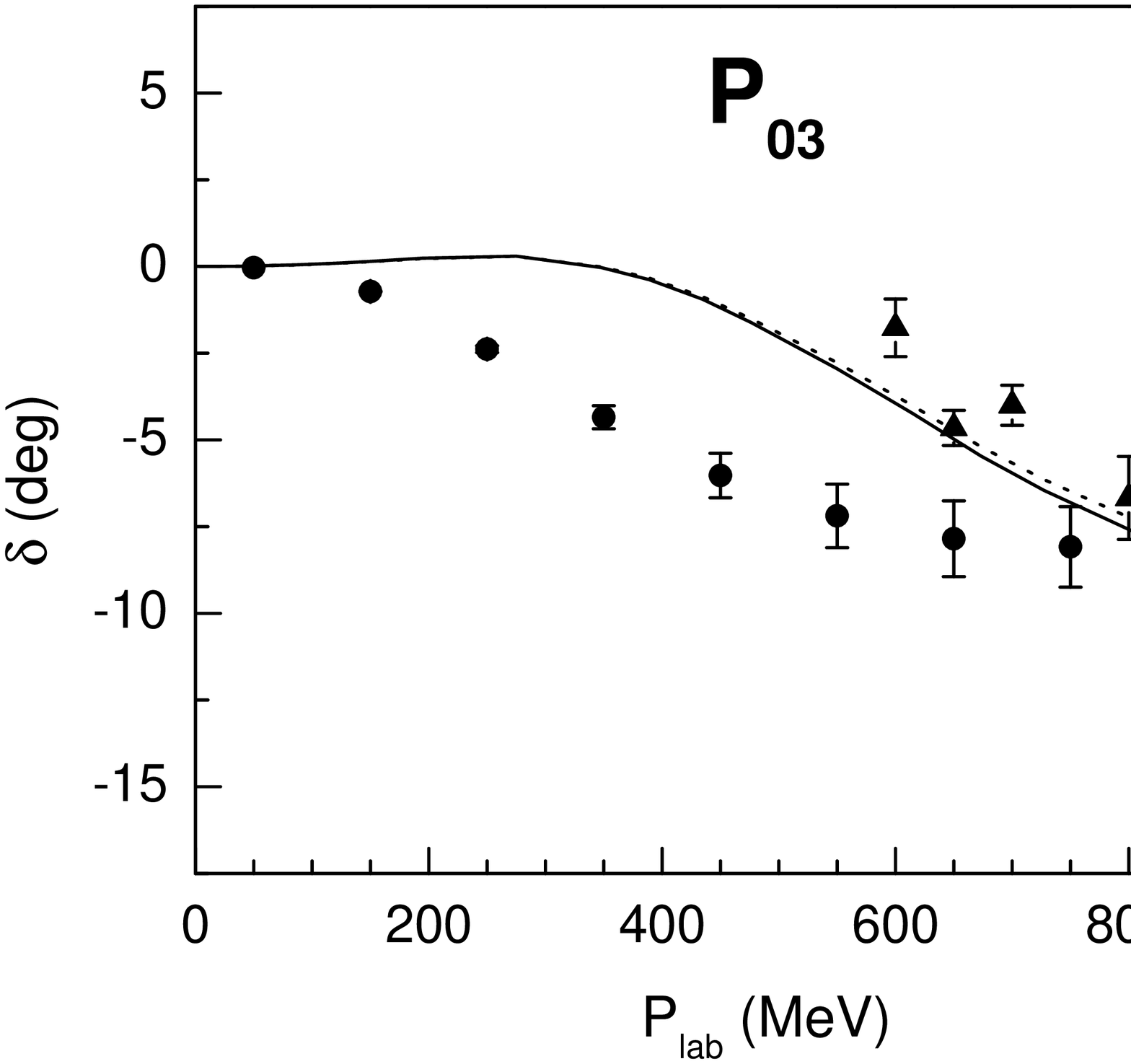,width=7.7cm}
\epsfig{file=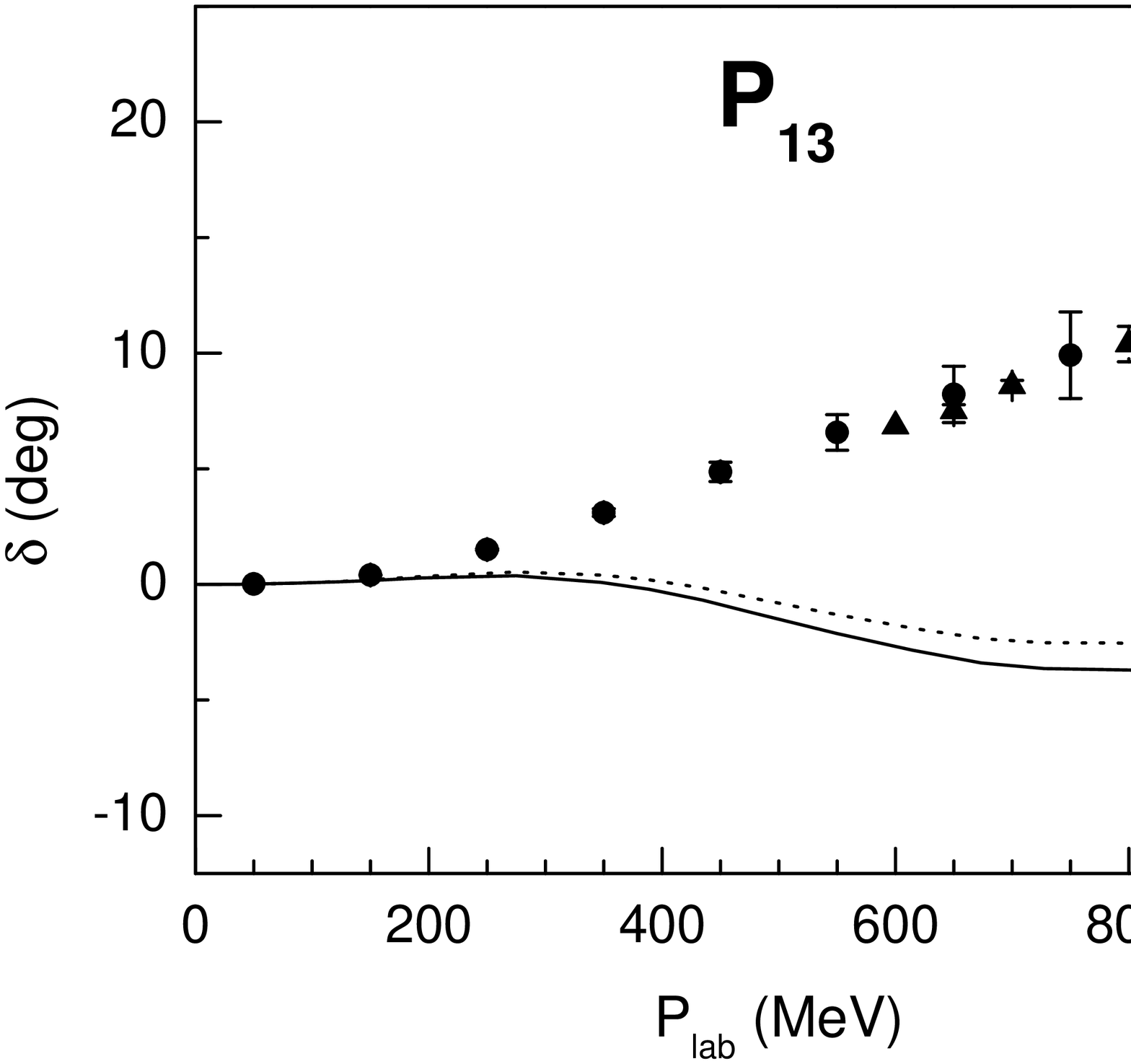,width=7.7cm} \vglue -2.3cm \caption{\small
\label{p0p1} $KN$ P-wave phase shifts. Same notation as in Fig.
\ref{s01s11}.}
\end{figure}

A RGM dynamical calculation is made to study the partial wave
phase shifts of $KN$ scattering by using the Hamiltonian, Eq.
$(\ref{hami5q})$, and the calculated  phase shifts of $S$ and $P$
waves with isospin $I=0$ and $I=1$ are shown in Figs. \ref{s01s11}
and \ref{p0p1} with solid lines.

For the $S_{01}$, $P_{01}$, $P_{11}$ and $P_{03}$ waves (here the
first subscript refers to the isospin quantum number $I$ and the
second one to the twice of the total angular momentum of the
system $2J$), our results are in agreement with the experimental
data. While for the $P_{13}$ channel our numerical phase shifts
are too repulsive when the laboratory momentum of the kaon meson
is greater than 500 MeV and $S_{11}$ channel a little repulsive.
Comparing with the results of the recent resonating group method
calculation of Lemaire {\it et al.} \cite{sle03} based on a
constituent quark model (CQM), in which the calculated phase
shifts of $S_{01}$, $P_{03}$, $P_{11}$ waves have opposite sign
and the $P_{01}$ channel is too repulsive for the experimental
data, we obtained the correct sign and reproduced the experimental
data quite well. This means that a reasonable interaction between
$K$ and $N$ can be obtained from the chiral SU(3) quark model when
the mixing of $\sigma_0$ and $\sigma_8$ mesons is considered as
ideal mixing and the mass of the $\sigma$ meson is taken to be 675
MeV, which is closely consistent with the relation
$m_{\sigma}=\sqrt{m_\pi^2+(2m_u)^2}$ from the dynamical vacuum
spontaneous breaking mechanism \cite{mds82}. We also compare our
results with those of the previous work of Black \cite{nbl02}.
Although our calculation achieves a considerable improvement on
the theoretical phase shifts in the magnitude for $S_{01}$,
$S_{11}$, $P_{01}$, $P_{11}$, $P_{03}$ waves, the results of the
$P_{13}$ channel are too repulsive in both Black's work and our
present one. Maybe the effects of the coupling to the inelastic
channels and hidden color channels should be considered in future
work.

Since there is something uncertain in the annihilation interaction
part, its influence on the phase shifts should be investigated. We
omitted the annihilation part entirely to see the effect, and
found that the numerical phase shifts only have very small changes.
This is because the annihilation part acts in the very short
range, so that it plays a nonsignificant role in the $KN$
scattering process.

One thing should be mentioned: in our present one channel
calculation for $KN$ scattering process the confinement potential
contributes pimping interactions between the two color singlet
clusters $K$ and $N$. Thus our numerical results will almost
remain unchanged; even the color quadratic confinement is replaced
by the color liner confinement or an improved one which is
presently unknown.

Recently we became aware of Ref. \cite{ybd04} written by Dai and
Wu, in which an investigation based on a dynamically spontaneous
symmetry breaking mechanism predicted that the mass of $\sigma$
meson is $m_\sigma=677$ MeV and the mixing angle between
$\sigma_0$ and $\sigma_8$ is $\theta^S=-18^\circ$. Using this
$m_\sigma$ and $\theta^S$, we calculated the $KN$ phase shifts and
the results are shown as dotted lines in Figs. \ref{s01s11} and
\ref{p0p1}. One can see that the $KN$ phase shifts can be also
explained quite well by taking this group of parameters. It is
comprehensible because in both of these two cases the attraction
of $\sigma$ is reduced, just in different approaches. When
$\theta^S=35.264^\circ$ (ideal mixing), the reduction comes from
the interaction between $u(d)$ and $s$ quarks vanished, while
$\theta^S=-18^\circ$, the interaction of $\sigma$ between two $u$,
$d$ quarks, is strongly reduced.

From the phase shifts of $KN$ (Figs. \ref{s01s11} and \ref{p0p1})
one can see that there is no signal for an existing $KN$ resonance
state both in $S$ and $P$ waves until the laboratory momentum of
the kaon meson stretches to 1 GeV. For studying the existence of
bound states of the $KN$ system, we solved the RGM equation for
the bound state problem [Eq. \ref{eqbound}]. The results showed
that the energies of the $KN$ system for both $S$ and $P$ waves
are located above the $KN$ threshold, which means that there is no
bound state. As a consequence, it can be said that the newly
observed exotic baryon $\Theta^+$ cannot be explained as a $KN$
resonance state or a $KN$ bound state in our present calculation.

\section{Conclusions}

The chiral SU(3) quark model is extended to the system with an
antiquark, and the $KN$ scattering process is studied by using
this model in the framework of the resonating group method. We
take the same initial input parameters as in our previous work,
which successfully explained the existing $NN$ and $YN$
experimental data. The difference is that in the present work the
mass of the scalar meson $\sigma$ is chosen to be 675 MeV (in our
precious work $m_\sigma=$595 or 625 MeV) and the mixing of
$\sigma_0$ and $\sigma_8$ is considered. Except for the case of
$P_{13}$, the numerical results of different partial waves are in
agreement with the experimental data. In comparison with the
previous results, our calculation achieves a considerable
improvement on the theoretical phase shifts. It seems that our
model can work well for the $KN$ system, in which an antiquark
$\bar{s}$ is there besides four $u(d)$ quarks, and the
interactions between two quarks obtained from this model might be
reasonable, which would be useful to study the structure of the
$\Theta^+$(1540) pentaquark state from the constituent quark model
point of view.

\begin{acknowledgements}
The authors wish to express their thanks to Dr. H.J. Wang for his
hospitality. He offered us Professor Dick Arndt's experimental
data, which were useful in our calculations. This work was
supported in part by the National Natural Science Foundation of
China No. 90103020.
\end{acknowledgements}

\end{document}